\documentclass[a4paper,11pt]{article}

\usepackage{jheppub} % for details on the use of the package, please
\title{\boldmath Signature of the Van der Waals like small-large charged AdS black hole phase transition in quasinormal modes }

\author{Yunqi Liu,}
\author{De-Cheng Zou,}
\author{and Bin Wang}
\affiliation{Department of Physics and Astronomy, Shanghai Jiao Tong
University, \\Shanghai 200240, China}

\emailAdd{liuyunqi@sjtu.edu.cn}
\emailAdd{zoudecheng@sjtu.edu.cn}
\emailAdd{wang$_-$b@sjtu.edu.cn}

\abstract{We calculate the quasinormal modes of massless scalar
perturbations around small and large four-dimensional
Reissner-Nordstrom-Anti de Sitter (RN-AdS) black holes. We find a
dramatic change in the slopes of quasinormal frequencies in small
and large black holes near the critical point where the Van der
Waals like thermodynamic phase transition happens. This further
supports that the quasinormal mode can be a dynamic probe of the
thermodynamic phase transition.}

 %\arxivnumber{1234.5678}

\begin{document}

\maketitle

\flushbottom

\section{Introduction}
\label{sec:intro}

Black hole thermodynamics has been an intriguing subject of
discussions for decades. Inspired by the AdS/CFT correspondence, the
black hole thermodynamics in the presence of a negative cosmological
constant has become even more appealing. The thermodynamic property
of AdS black holes was first investigated in \cite{HawkingPage83}
where it was found that there exists a Hawking-Page phase transition
between the Schwarzschild AdS black hole and pure AdS space. Later
it was further disclosed that in the charged AdS black holes there
is a first order phase transition between small and large black
holes in the canonical ensemble \cite{Chamblin99,Chamblin992}. This
phase transition was argued superficially analogous to a liquid-gas
phase transition in the Van der Waals fluid. The superficial
reminiscence was also observed in other AdS black hole backgrounds
\cite{Niu12,Fernando06,Dey07,Dey072,Anninos,Poshteh,WeiLiu,Lala,Tsai,Banerjee1,Banerjee2,Banerjee3,Banerjee4,Banerjee11}.
Recently the study of thermodynamics in AdS black holes has been
generalized to the extended phase space, where the cosmological
constant is identified with thermodynamic pressure, $
P=-\frac{\Lambda}{8 \pi}=\frac{3}{8\pi l^2}$, in the geometric units
$G_N=\hbar=c=k=1$. $l$ denotes the AdS radius, which is considered
varying \cite{Dolan,Dolan2,Dolan3,DolanKastor,Spallucci} and is
included in the first law of black hole thermodynamics to ensure the
consistency between the first law of black hole thermodynamics and
the Smarr formula \cite{Kastor:2009wy}. With the varying
cosmological constant, the AdS black hole mass is identified with
enthalpy and there exists a natural conjugate thermodynamic volume
to the cosmological constant. In the extended phase space with
cosmological constant and volume as thermodynamic variables, it was
interestingly observed that the small-large black hole system admits
a more direct and precise coincidence with the Van der Waals system
\cite{Kubiznak}. More discussions on comparing phase transitions in
AdS black holes with the Van der Waals analogy can be found in
\cite{Gunasekaran,BelhajChabab,BelhajChabab1,Hendi,Belhaj,ChenLiu,CaiCao,XuXu,Dutta,Zhao,
Altamirano:2013ane,Altamirano:2013uqa,Zou:2013owa,Mo:2014mba,Zou:2014mha}.

It has been an expectation for a long time that
black hole thermodynamical phase transitions can
have some observational signatures to be
detected. Considering that quasinormal modes of
dynamical perturbations  are characteristic
sounds of black holes
\cite{Nollert,Kokkotas,WangBraz,KonoplyaZhidenko},
it is expected that black hole phase transitions
can be reflected in the dynamical perturbations
in the surrounding geometries of black holes
through frequencies and damping times of the
oscillations.

In asymptotically flat Reissner-Nordstrom (RN) black hole, it was
argued that a second order phase transition happens where the heat
capacity appears singular \cite{Davies,Davies2}. This
thermodynamical second order phase transition point was disclosed in
the dynamical quasinormal modes \cite{Jingpan08}. The observed
relation between thermodynamical phase transitions and dynamical
perturbations is not trivial \cite{BertiCardoso} and was also
confirmed in \cite{He:2008im}. In AdS black holes, thermodynamic
phase transition in the dual field theory corresponds to the onset
of instability of a black hole, so that quasinormal modes of black
holes are naturally connected with thermodynamic phase transitions
of strongly coupled field theories \cite{Gubser:2000mm}. Besides,
the second-order phase transition of a topological AdS black hole to
a hairy configuration was found reflected in the quasinormal modes
of the electromagnetic and scalar perturbations, respectively
\cite{KoutsoumbasMusiri,ShenWang}. Moreover the phase transition in
the charged topological-AdS black holes was observed in the
quasinormal modes of the electromagnetic and gravitational
perturbations \cite{Koutsoumbas:2008pw}. Different phase properties
between the massless BTZ black hole and the generic nonrotating BTZ
hole were also detected in the scalar field as well as the fermonic
field perturbations \cite{Rao:2007zzb}. Phase transition between
scalar and non-rotating BTZ black holes in three dimensions was also
shown in the dynamical perturbation behaviors \cite{Myung:2008ze}.
In \cite{Gubser:2000ec}, it was even argued that  the
thermodynamical stability is closely related to the dynamical
stability for black brane solutions. More recently, the phase
transitions before and after the scalar field condensation in the
backgrounds of the AdS black hole and AdS soliton were further
observed in the quasinormal modes of dynamical perturbations
\cite{He:2010zb, Cai:2011qm,Liu:2011cu,Abdalla:2010nq}.

It is of great interest to generalize the discussions on the
relation between thermodynamical phase transitions and dynamical
perturbations to the Van der Waals like phase transitions in RN-AdS
black holes. In this paper we will concentrate on this topic and
disclose the fact that quasinormal modes can again be a probe of the
phase transition in the Van der Waals analogy system. In Sec.II we
will first review the analogy of the small-large black hole system
with the Van der Waals system in the extended phase space. In Sec.
III, we will disclose numerically that different phases can be
reflected by the quasinormal modes of dynamical perturbations. We
will summarize our results in the last section.

\section{Phase transition in charged AdS black hole spacetime}
\label{2s}

We consider a four-dimensional RN-AdS black hole
with the metric
\begin{eqnarray}
&&ds^2=-f(r)dt^2+\frac{1}{f(r)}dr^2+r^2d\Omega^2_2,\label{eq:2a}\\
&&f(r)=1+\frac{r^2}{l^2}-\frac{2M}{r}+\frac{Q^2}{r^2},\label{eq:3a}
\end{eqnarray}
where $M$ and $Q$ are the mass and charge of the
black hole. In terms of the black hole event
horizon $r_H$, the mass $M$, Hawking temperature
$T$, entropy $S$ and electromagnetic potential
$\Phi$ of the RN-AdS black hole can be expressed
as
\begin{eqnarray}
M&=&\frac{r_H}{2}\left(1+\frac{r_H^2}{l^2}+\frac{Q^2}{r_H^2}\right),\quad
T=\frac{1}{4\pi r_H}\left(1+\frac{3r_H^2}{l^2}-\frac{Q^2}{r_H^2}\right),\nonumber\\
S&=&\pi r_H^2, \quad \Phi=\frac{Q}{r_H}.\label{eq:4a}
\end{eqnarray}

Considering the thermodynamic volume $V=4\pi r_H^3/3$ and the
pressure $P=3/(8\pi l^2)$, we can have the first law of black hole
thermodynamics in an extended phase space \cite{Kubiznak}
\begin{eqnarray}
dM=TdS+\Phi dQ+VdP,\label{eq:5a}
\end{eqnarray}
where the black hole mass $M$ can be considered as the enthalpy
rather than the internal energy of the gravitational system
\cite{Kastor:2009wy}.

In addition, the expression of the black hole temperature can be
translated into the equation of state  $P(V,T)$
\begin{eqnarray}
P=\frac{T}{2r_H}-\frac{1}{8\pi
r_H^2}+\frac{Q^2}{8\pi r_H^4}, \quad
r_H=(3V/4\pi)^{1/3}.\label{eq:6a}
\end{eqnarray}

Then the critical point can be obtained from
\begin{eqnarray}
\frac{\partial P}{\partial r_H}\Big|_{T=T_c, r_H=r_c}
=\frac{\partial^2 P}{\partial r_H^2}\Big|_{T=T_c, r_H=r_c}=0,
\end{eqnarray}
which leads to $T_c=\frac{\sqrt{6}}{18\pi Q }$,
$r_c=\sqrt{6}Q$ and $P_c=\frac{1}{96\pi Q^2}$.

In the extended phase space, the Gibbs free energy for fixed charge
is read
\begin{eqnarray}
G(T,P)&=&\frac{1}{4}\left[r_H-\frac{8\pi P
r_H^{3}}{3}-\frac{3Q^2}{r_H}\right].\label{eq:7a}
\end{eqnarray}
Here $G$ is understood as a function of pressure and temperature by
considering the equation of state Eq.(\ref{eq:6a}).

Fig.\ref{PT0} shows the Gibbs free energy and temperature coincide
for small and large black holes \cite{Kubiznak}. The curve is the
coexistence line of small-large black hole phase transition of the
charged AdS black hole system. The critical point is highlighted by
a small circle at the end of the coexistence line. From Fig.1 we can
see that to accommodate the phase transition from small to large
black holes, we can choose either the isobaric or isothermal
process.  Then, we display the behaviors of the Gibbs free energy in
the isobaric process by fixing $P$ in Fig.\ref{freenergyt0} and the
Gibbs free energy in the isothermal process by fixing the
temperature of the system in Fig.\ref{freenergyp0}, respectively. It
is clear that both $G$ surfaces demonstrate the characteristic
swallow tail behavior, marking the first order transition in the
system. In addition, the corresponding $T-r_H$ and $P-r_H$ diagrams
of charged AdS black holes for the isobaric and isothermal processes
are also shown in Fig.\ref{freenergyt0} and Fig.\ref{freenergyp0},
respectively. Both of them contain inflection points and the
behaviors are reminiscent of Van der Waals system \cite{Kubiznak}.

\begin{figure}[ht]\label{PT0}
\centering
\includegraphics[width=140pt,angle=270]{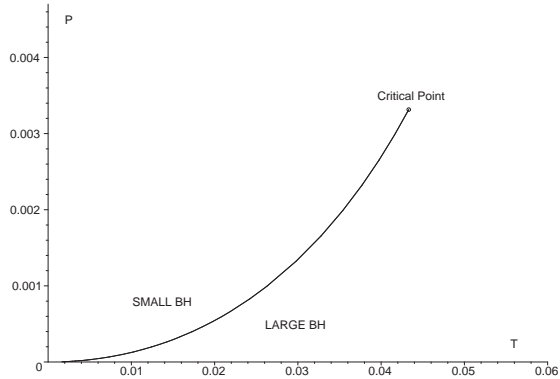}
\caption{\label{PT0} The coexistence line of charged AdS black hole
of small-large black hole phase transition in P-T plane with a unit
black hole charge.}
\end{figure}

\section{Perturbations in the charged AdS black hole
spacetime}

Now we consider a massless scalar field
perturbation in the surrounding geometry of the
four-dimensional RN-AdS black hole background
Eq.(\ref{eq:2a}). The radial part of the
perturbation, $\Psi(r,t)=\psi(r) e^{-i \omega
t}$, is described by the Klein-Gordon equation
\begin{eqnarray}\label{psieqn}
\psi''(r)+\left[\frac{f'(r)}{f(r)}+\frac{2}{r}\right] \psi'(r)+\frac{w^2 \psi(r)}{f(r)^2}=0,
\end{eqnarray}
where $\omega$ indicates the frequency of the
perturbation.

Near the black hole horizon $r_H$, we can impose
the ingoing boundary condition,
$\psi(r)\sim(r-r_H)^{-i \frac{\omega}{4\pi T}}$.
Defining $\psi(r)$ as $\psi(r) exp[-i\int
\frac{\omega}{f(r)}dr]$, where $exp[-i\int
\frac{\omega}{f(r)}dr]$ asymptotically approaches
the ingoing wave near horizon, we can rewrite
Eq.(\ref{psieqn}) into
\begin{eqnarray}\label{psieqn0}
\psi''(r)+\psi'(r) \left[\frac{f'(r)}{f(r)}-\frac{2 i \omega}{f(r)}+\frac{2}{r}\right]- \psi(r)\frac{2 i \omega}{r f(r)}=0,
\end{eqnarray}
so that when $r\rightarrow r_H$, we can set $\psi(r)=1$.  At the AdS
boundary $r\rightarrow \infty$, we need $\psi(r)=0$. With the
boundary conditions we solve  Eq.(\ref{psieqn0}) numerically and
find the frequencies of the quasinormal modes  by using  the
shooting method.

We are going to study whether the signature of
thermodynamical first order phase transition in
charged AdS black holes, in analogy to the
liquid-gas phase transition, can be reflected in
the dynamical quasinormal modes behavior in the
massless scalar perturbation. We will examine the
dynamical perturbations in the possible two
processes, namely isobaric process by fixing the
pressure $P$ and the isothermal process by fixing
the temperature $T$, to accommodate the phase
transitions between small-large black holes. We
will set the black hole charge to be unity in our
following numerical computations.

\subsection{Isobaric phase transition}
In this case, $P$ (or $l$) is fixed so that the black hole horizon
$r_H$ is the only variable in the system.  We have seen the behavior
of isobar in the $T-r_H$ diagram in Fig.\ref{freenergyt0}. The
oscillating part is for $P<P_c$, where a small-large black hole
phase transition occurs in the system. The critical isobar $P_c$ is
got by $\frac{\partial T}{\partial r_H}=\frac{\partial^2 T}{\partial
r_H^2}=0$. More information of the phase transition is reflected in
the Gibbs free energy in the left panel of Fig.\ref{freenergyt0}.
The solid line and the dotted line cross each other at the
intersection point marked as ``5", which indicates the coexistence
of two phases in equilibrium. In the right panel this point is
separated into ``L5" and ``R5", which have the same Gibbs free
energy and the same black hole temperature $T=T_c\simeq 0.02630$ for
small and large black holes. Combining the Gibbs free energy and the
phase diagram, we find that the physical phase marked between points
``1-5" or ``1-L5" is for the small black hole, while physical phase
indicated between points ``5-4" or ``R5-4" is for the large black
hole.

\begin{figure}[ht]\label{freenergyt0}
\centering
\includegraphics[width=180pt]{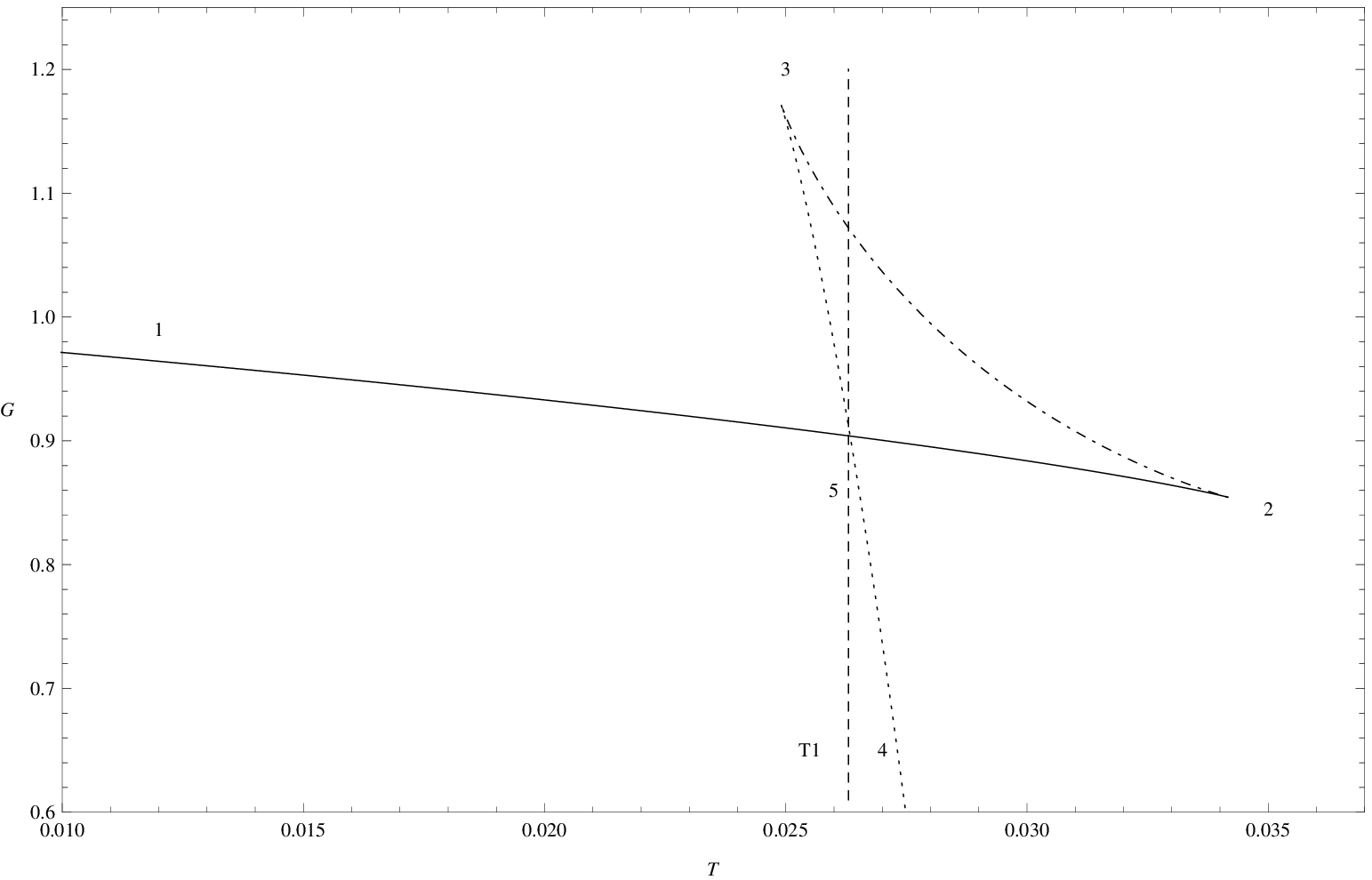}
\includegraphics[width=180pt]{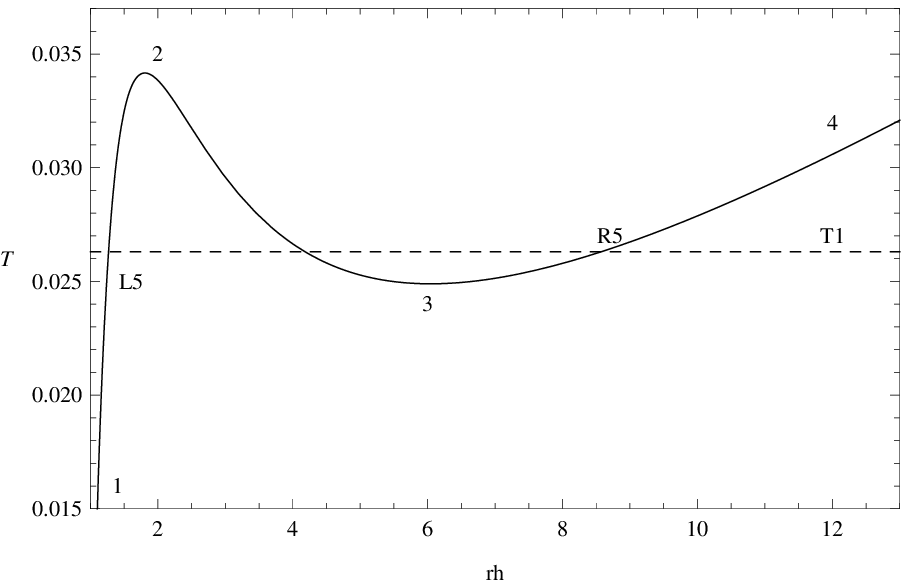}
\caption{\label{freenergyt0} Left) The Gibbs free energy is given as
a function of black hole temperature for $P=0.001$. Solid
line(Line1-2) and dotted line(Line3-4) cross at the point ``5",
indicating the place where the phase transition happens. Right) The
black hole temperature $T$ is depicted as a function of the black
hole horizon $r_H$ for $P=0.001$. The dashed line indicates the
phase transition temperature. }
\end{figure}

\begin{figure}[ht]\label{freenergyp0}
\centering
\includegraphics[width=180pt]{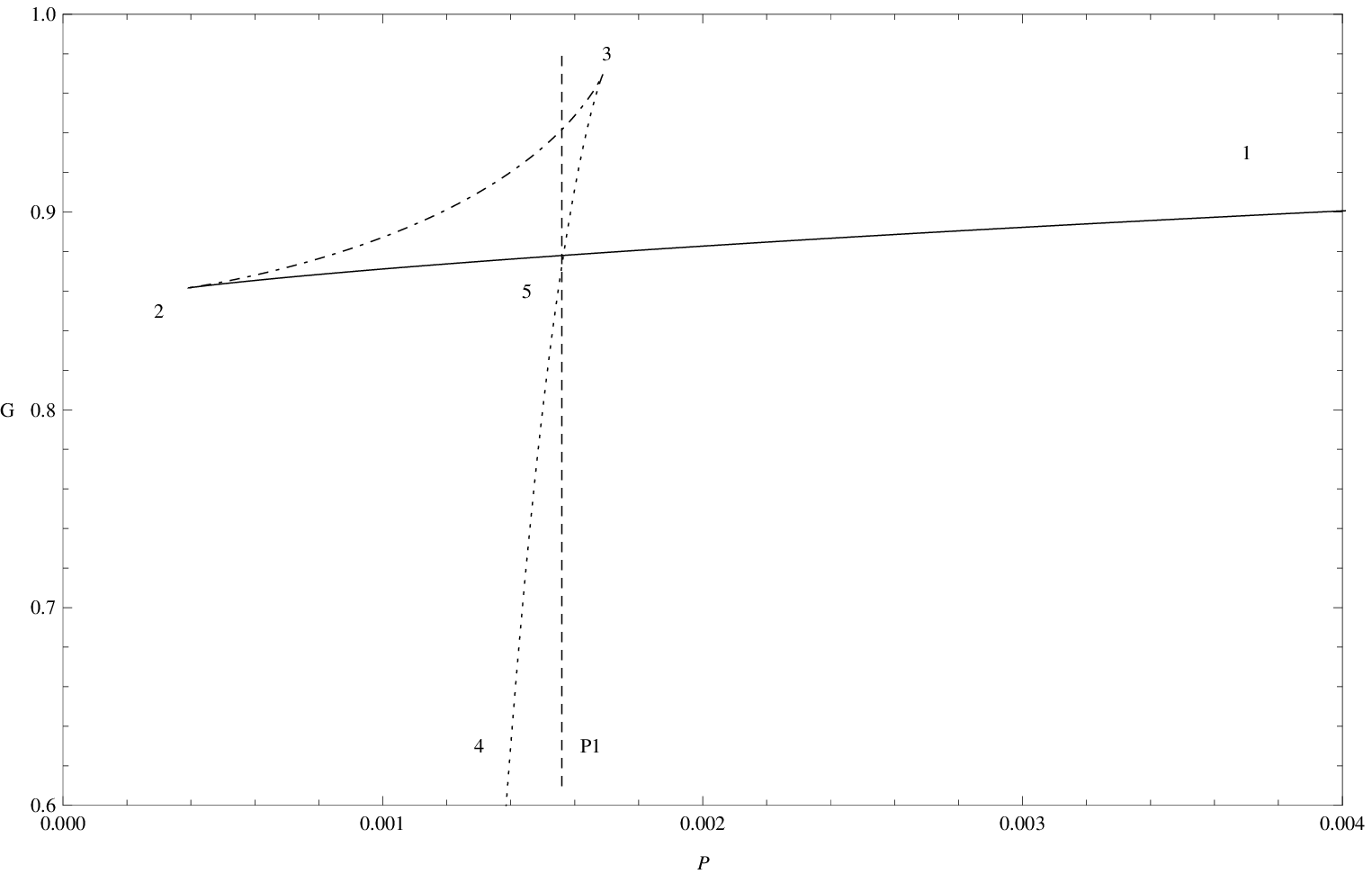}
\includegraphics[width=180pt]{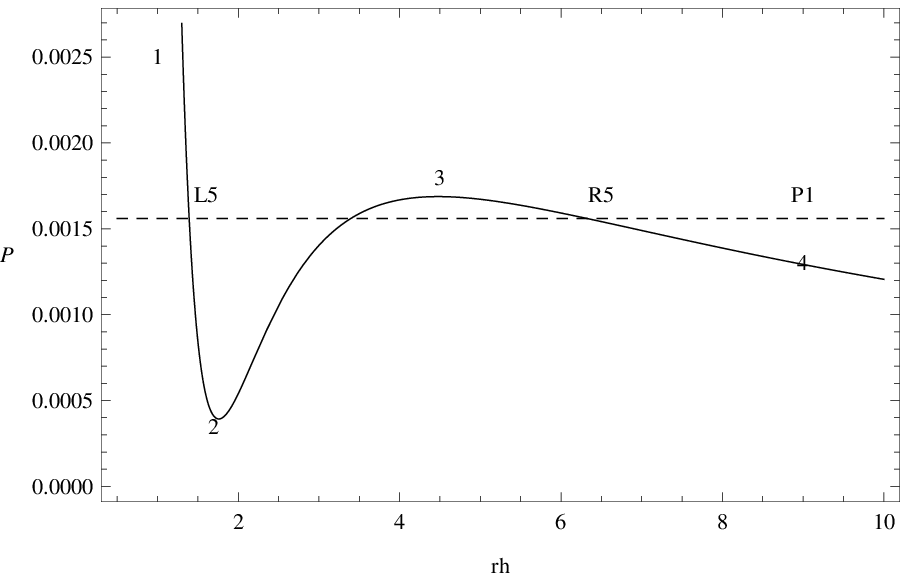}
\caption{\label{freenergyp0}  Left) The Gibbs free energy is given
as a function of pressure $P$ for the isothermal phase transition
with $T=0.0032$. Solid line 1-2 and dotted line 3-4 cross each other
at the point ``5", which indicates the phase transition point.
Right) The black hole pressure $P$ is depicted as a function of
black hole horizon $r_H$ with $T=0.0032$. The dashed line indicates
the phase transition pressure.}
\end{figure}

In Table \ref{table1}, we list the frequencies of the quasinormal
modes of the massless scalar perturbation around  small and large
black holes described between points ``1-L5" and ``R5-4"
respectively in Fig.\ref{freenergyt0} for different temperatures.
The frequencies above the horizontal line are for the small black
hole phase while the frequencies below are for the large black hole
phase.
\begin{center}
\begin{table}[ht]\label{table1}
\centering
\begin{tabular}{|c|c|c|}
\hline
$T(10^{-2})$ & $r_H$ & $\omega$  \\[0.8ex] \hline
2.2& 1.190&0.2223-0.02397I \\
2.3& 1.206&0.2222-0.02408I \\
2.4&1.223&0.2222-0.02421I  \\
2.5&1.242&0.2221-0.02436I  \\
2.55& 1.252&0.2221-0.02445I \\
2.6& 1.262&0.2220-0.02454I \\ \hline
2.65&8.774 &0.2364-0.1963I \\
2.7& 9.248 &0.2403-0.2070I \\
2.8&10.099&0.2478-0.2262I \\
2.9& 10.871 &0.2551-0.2436I \\
3.1& 12.282 &0.2695-0.2752I \\
\hline
\end{tabular}
\caption{\label{table1}The quasinormal frequencies of massless
scalar perturbation with the change of black hole temperature. The
upper part, above the horizontal line, is for the small black hole
phase, while the lower part is for the large black hole phase. }
\end{table}
\end{center}

For the small black hole phase, we see that when
the temperature decreases from the phase
transition critical point $T_c$, the black hole
becomes smaller. In this process the real part
frequencies change very little, while the
absolute values of the imaginary part of
quasinormal frequencies decrease. This result is
consistent with the objective picture obtained in
\cite{zhuWang}. Considering that the decay of the
test field outside the black hole is due to the
black hole absorption, it is natural to
understand that when the black hole becomes much
smaller, its absorption ability decreases so that
the field will decay slower and the oscillation
(real part of the perturbation) nearly keeps as a
constant.

For the large black hole phase, we find that when the temperature
increases from the critical value $T_c$, the black hole gets bigger.
The real part together with the absolute value of the imaginary part
of quasinormal frequencies increase. The massless scalar
perturbation outside the black hole gets more  oscillations but it
decays faster. To have a physical picture of this result, we plot
the effective potential behavior near the AdS boundary. In the AdS
boundary, when $r\rightarrow \infty$, the tortoise coordinate
$r^*_{AdS}$ tends to a constant value \cite{WangMolina}. With the
increase of $r_H$, $r^*_{AdS}$ at the AdS boundary becomes smaller.
Thus the infinite potential wall at the AdS boundary will be moved
towards the black hole. This will bounce back more outgoing
perturbation towards the black hole, which will add energy to the
perturbation and make the real part frequency increase. On the other
hand, when the black hole becomes bigger, it becomes more greedy and
can absorb more things. This can explain that the absolute value of
the imaginary part increases with the black hole size, so that the
decay of the perturbation becomes faster.

The drastically different quasinormal frequencies for small and
large black hole phases are plotted in Fig.\ref{arrow12}. In the
left panel, to plot the figure we took the value of the real part as
five significant digits while it was taken as four in Table
\ref{table1}, actually the real part varies little here. From the
figure, we can see different slopes of the quasinormal frequencies
in the massless scalar perturbations reveal that small and large
black holes are in different phases. To change the small black hole
to be a large one, one has to encounter a phase transition. The
different properties disclosed here in dynamical perturbations
reflect the idea of the thermodynamic phase transitions between
small-large RN-AdS black holes.

\begin{figure}[ht]\label{arrow12}
\centering
\includegraphics[width=180pt]{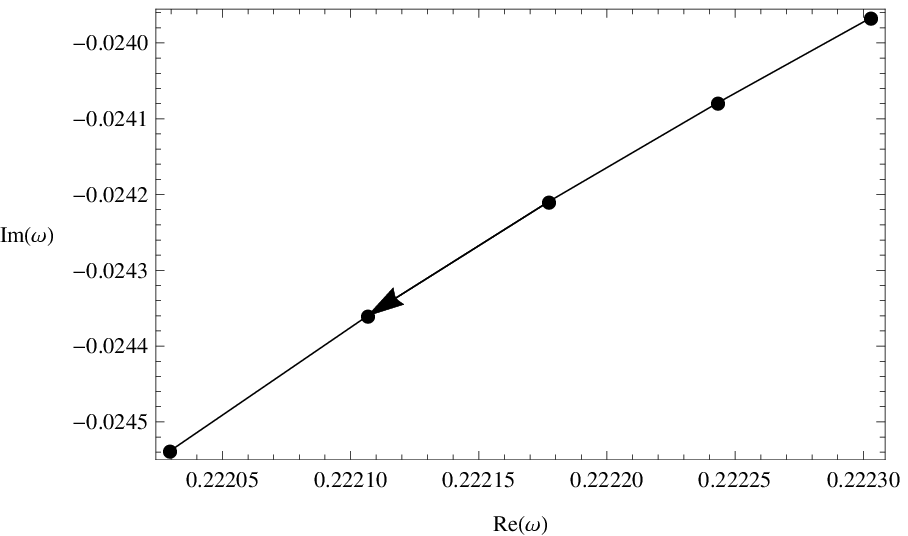}
\includegraphics[width=180pt]{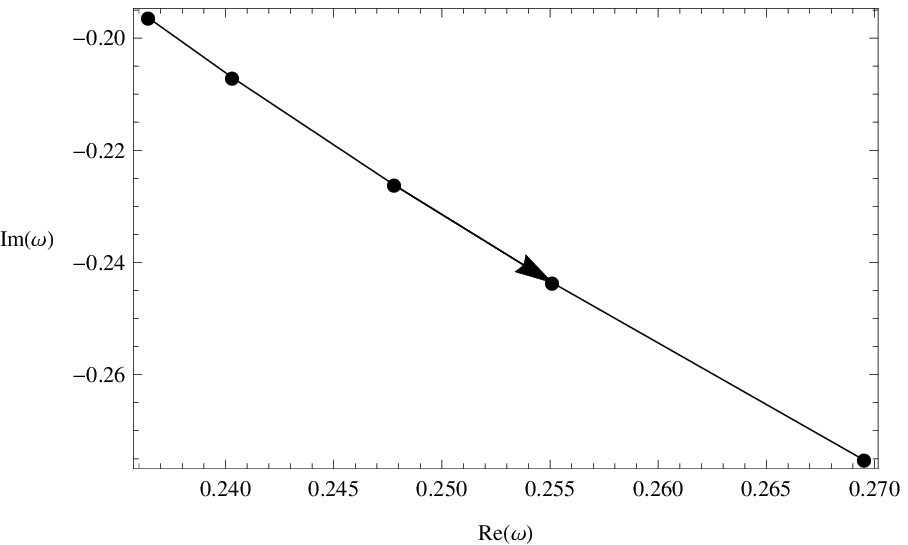}
\caption{\label{arrow12} The left panel shows the
behavior of quasinormal modes for small black
holes and the right panel shows the behavior of
quasinormal modes for large black holes. The
arrow indicates the increase of black hole
horizon. }
\end{figure}

\subsection{Isothermal phase transition}
Fixing the black hole temperature $T$, we can plot the $P-r_H$
diagram of charged AdS black holes in the right panel of
Fig.\ref{freenergyp0}. For $T<T_c$ there is an inflection point and
the behavior is reminiscent of the Van der Waals system. The
critical point can be got from $\frac{\partial P}{\partial
r_H}=\frac{\partial^2 P}{\partial r_H^2}=0$. The behavior of the
Gibbs free energy is plotted in the left panel of Fig.3. The cross
point ``5" between the solid line marked as ``1-5" and the dotted
line denoted as ``4-5" shows that the Gibbs free energy and $P$
coincide for small and large black holes. In the right panel of the
phase diagram, the point ``5" is separated into ``L5" and ``R5" for
the same Gibbs free energy and the chosen $P\simeq 0.00156$ where
the small and large black hole can coexist. Combining the Gibbs free
energy and the phase diagram, we find that the physical phase marked
between points ``1-5" or ``1-L5" is for the small black hole, while
physical phase indicated between points ``5-4" or ``R5-4" is for the
large black hole.

\begin{center}
\begin{table}[ht]\label{table2}
\centering
\begin{tabular}{|c|c|c|}
\hline
$P(10^{-3})$ & $r_{H}$ & $\omega$ \\[0.8ex] \hline
2.0&1.350&0.2955-0.06489I  \\
1.9& 1.359&0.2892-0.06100I  \\
1.8&1.368&0.2827-0.05714I  \\
1.7& 1.377&0.2761-0.05328I  \\
1.6&1.387&0.2692-0.04944I \\ \hline
1.5&6.906 &0.2860-0.2320I \\
1.45&7.389 &0.2854-0.2400I \\
1.4&7.880 &0.2848-0.2472I \\
1.35&8.388 &0.2840-0.2538I \\
1.3&8.919 &0.2831-0.2599I \\
1.25&9.480 &0.2822-0.2656I\\
1.2&10.075 &0.2811-0.2710I\\
1.1&11.396 &0.2790-0.2809I\\
\hline
\end{tabular}
\caption{\label{table2}The quasinormal
frequencies of massless scalar perturbations for
black holes with different sizes in the
isothermal phase transition with $T=0.032$. The
upper part, above the horizontal line, is the
frequency for the small black hole phase, while
the lower part is for the large black hole
phase.}
\end{table}
\end{center}

In  Table \ref{table2} we list the frequencies of quasinormal modes
for small and large black hole phases in the isothermal phase
transition with $T=0.032$. The data above the horizontal line are
the frequencies for small black holes, while those below are for
large black holes. We can see that for the small black hole phase,
as the black hole horizon grows, the corresponding pressure $P$
decreases. In this process the real parts of the frequencies
decrease and the absolute imaginary parts decrease as well. For the
large black hole case, with the increase of the black hole size, the
pressure decreases.  In this process, the real parts of the
quasinormal frequencies decrease while the absolute values of the
imaginary parts increase.   Fig.\ref{arrow34} shows the behaviors of
the quasinormal modes for small and large holes. The arrows indicate
the increase of the black hole size. It is clear that in small and
large black hole phases, the properties of the  quasinormal
frequencies are completely different.

\begin{figure}[ht]\label{arrow34}
\centering
\includegraphics[width=180pt]{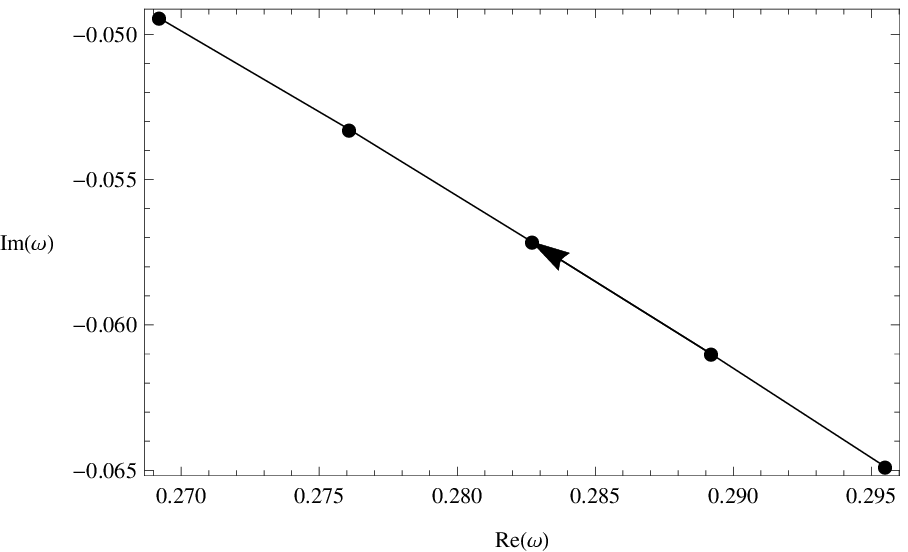}
\includegraphics[width=180pt]{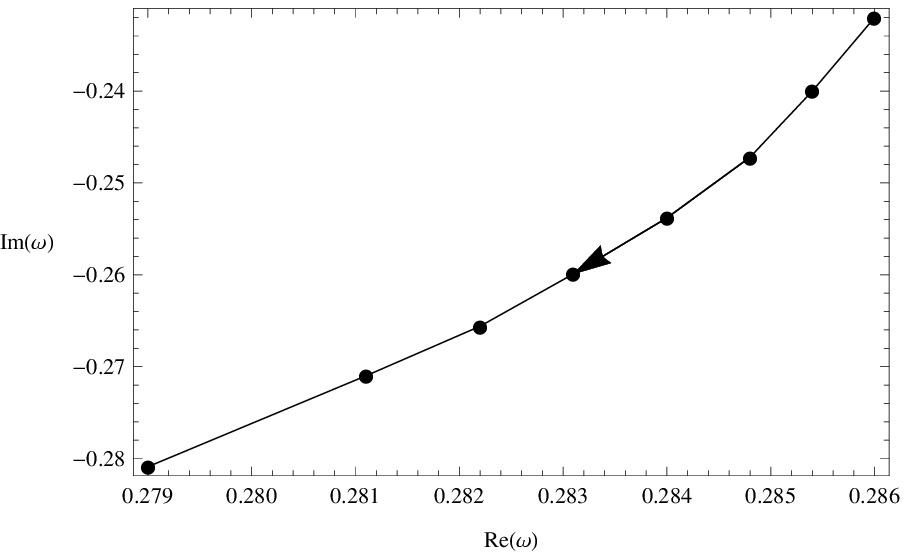}
\caption{\label{arrow34} The left panel shows the
behavior of quasinormal modes for small black
holes and the right panel exhibits the behavior
of  quasinormal modes for large black holes. The
arrows in the figure indicate the increase of the
black hole size.}
\end{figure}

In the isothermal phase transition, besides the variable $r_H$, the
AdS radius $l$ is another variable to influence the quasinormal
behaviors. They are related by the fixed temperature, however each
of them has independent influence on the quasinormal modes. Now
let's first examine the independent influence on the quasinormal
frequencies by each of these two parameters. By fixing $l$ ($P$), we
list the influence of $r_H$ on the frequencies for small and large
black holes in Table \ref{table3} and \ref{table31}. As the physical
interpretation given above, for the small black hole phase the
quasinormal modes exhibits a nearly constant oscillation and decays
slower when the black hole further shrinks. For the large black
holes when $r_H$ increases, as shown in Fig.\ref{potential0} the AdS
wall moves closer to the black hole which will bounce back the
outgoing perturbations more strongly to increase the real part of
the quasinormal frequency. At the meantime, the black hole becomes
bigger to swallow more things, which results in the fast decay of
the perturbation.

Now let's turn to discuss the independent effect
of $l$ ($P$) on the quasinormal frequency by
fixing the black hole size $r_H$. For small and
large black hole phases, the results are listed
in Table \ref{table4} and \ref{table41}. For both
small and large holes, with the decrease of $P$
(the increase of $l$), both the real parts and
the absolute imaginary parts of quasinormal
frequencies decrease.  The effects of increasing
$l$ on the quasinormal frequency for small and
large holes are completely different from the
effect of increasing the black hole size $r_H$.
This can be attributed to different boundary
behaviors caused by the increases of $l$ and
$r_H$. With the increase of $r_H$, we saw in
Fig.\ref{potential0} that the AdS wall moves
closer to the black hole. But when the AdS radius
$l$ increases, the AdS wall moves further away,
which is illustrated in Fig.\ref{potential}. This
results in the mildly bouncing back of the
outgoing perturbations when $l$ increases
compared with the increase of $r_H$, which
explains the decrease of the real part of the
frequency.  But the outgoing perturbation will be
mildly bounced back continuously with the
increase of $l$, while the black hole size is
fixed so that the amount of perturbation it can
swallow is fixed. This accounts for the physical
reason that the decay becomes slower when $l$
increases.

Thus we see that the influences given by $r_H$
and $l$ on quasinormal frequencies of the
perturbation are different. The competition
between these two factors result in the
properties of the quasinormal frequencies for
small and large black holes shown in Table
\ref{table2} and Fig.\ref{arrow34}.

To see closely how these two factors compete with
each other to influence the quasinormal
frequencies, we do a double-series expansion of
the frequency $\omega(r_H+\triangle r_H,
P+\triangle P)$ in $(\triangle r_H,\triangle P),$
\begin{eqnarray}\label{series}
&&\omega(r_H+\triangle r_H, P+\triangle P)\nonumber\\
&&=\omega(r_H,P)+\partial_{r_H} \omega \triangle r_H+ \partial_{P}
\omega \triangle P+O(\triangle r_H ^2,\triangle P^2,\triangle r_H
\triangle P ).
\end{eqnarray}
This means that the changes of the quasinormal
frequency get  two influences, one is from the
change of the black hole size $r_H$, and the
other  is from the change of the pressure $P$ (or
AdS radius $l$). For simple discussions in the
following, we define $\Delta_1\equiv
\partial_{r_H} \omega \triangle r_H $ and
$\Delta_2\equiv \partial_{P} \omega \triangle P$.

\begin{center}
\begin{table}[ht]\label{table3}
\centering
\begin{tabular}{|c|c|c|}
\hline
$P(10^{-3})$&$r_H$ & $\omega$ \\[0.8ex] \hline
2.0&1.35 & 0.29547-0.06489I\\[0.8ex] \hline
2.0&1.40 & 0.29542-0.06604I\\[0.8ex] \hline
2.0&1.45 & 0.29540-0.06734I\\[0.8ex] \hline
2.0&1.50 & 0.29540-0.06875I\\[0.8ex] \hline
\end{tabular}
\caption{\label{table3} This table shows how the
quasinormal frequencies change as the black hole
horizons $r_H$ increase for small black holes.
The pressure is fixed as $P=0.002$.}
\end{table}
\end{center}
\begin{center}
\begin{table}[ht]\label{table31}
\centering
\begin{tabular}{|c|c|c|}
\hline
$P(10^{-3})$&$r_H$ & $\omega$\\[0.8ex] \hline
1.4& 7.70& 0.2826-0.2415I \\[0.8ex] \hline
1.4& 7.88& 0.2848-0.2472I\\[0.8ex] \hline
1.4& 8.00& 0.2862-0.2510I\\[0.8ex] \hline
\end{tabular}
\caption{\label{table31}This table shows how the
quasinormal frequencies change as black hole
horizons $r_H$ increase for large black holes.
The pressure is fixed as $P=0.0012$.}
\end{table}
\end{center}

\begin{center}
\begin{table}[ht]\label{table4}
\centering
\begin{tabular}{|c|c|c|}
\hline
$r_H$ &$P(10^{-3})$ & $\omega$\\[0.8ex] \hline
1.35& 1.8& 0.2827-0.0578I\\[0.8ex] \hline
1.35& 1.9& 0.2892-0.0608I\\[0.8ex] \hline
1.35& 2.0& 0.2955-0.0649I\\[0.8ex] \hline
1.35& 2.1& 0.3016-0.0690I\\[0.9ex] \hline
\end{tabular}
\caption{\label{table4}This table shows how the
quasinormal frequencies change as the pressure
$P$ increases for small black holes, where the
black hole horizon is fixed as $r_H=1.4$.}
\end{table}
\end{center}
\begin{center}
\begin{table}[ht]\label{table41}
\centering
\begin{tabular}{|c|c|c|}
\hline
$r_H$ &$P$ & $\omega$ \\[0.8ex] \hline
7.880&0.0013 & 0.2712-0.2294I \\[0.8ex] \hline
7.880&0.0014 & 0.2848-0.2472I\\[0.8ex] \hline
7.880&0.0015 & 0.2982-0.2650I\\[0.8ex] \hline
\end{tabular}
\caption{\label{table41}This table shows how the
quasinormal frequencies change as the pressure
$P$ increases for large black holes, where the
black hole horizon is fixed as $r_H=11$.}
\end{table}
\end{center}
\begin{figure}[ht]\label{potential0}
\centering
\includegraphics[width=180pt]{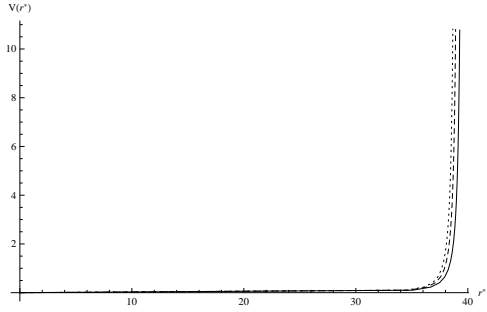}
\caption{\label{potential0} The potential is
depicted for the fixed pressure $P=0.0014$ (or
the fixed AdS radius $l$).  At the AdS boundary,
$r^*$ approaches a constant. With the increase of
the black hole size (the solid line is for
$r_H=7.7$, the dashed is for $r_H=7.88$ and the
dotted line for $r_H=8$), the constant of $r^*$
decreases, so that the potential wall moves
towards the black hole. }
\end{figure}
\begin{figure}[ht]\label{potential}
\centering
\includegraphics[width=180pt]{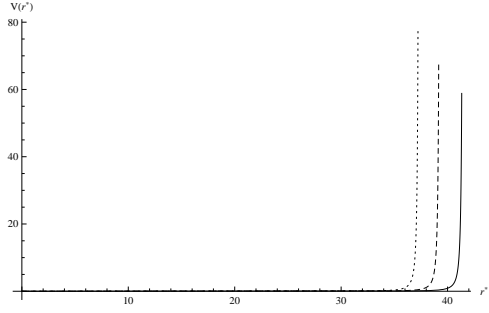}
\caption{\label{potential} The potential is depicted for the fixed
black hole size $r_H=7.88$, but different pressure $P$(or AdS radius
$l$). At the AdS boundary, $r^*$ approaches a constant. With the
increase of the pressure $P$(or the decrease of AdS radius $l$), the
potential wall moves towards the black hole. The solid line is for
$P=0.0013$, the dashed line is for $P=0.0014$ and the dotted for
$P=0.0015$. }
\end{figure}

Considering the equation of state Eq.(\ref{eq:6a}), the changes of
$P$ and $r_H$ are not completely independent. They are related by
\begin{eqnarray}\label{totaldev}
dP=(\frac{1}{4\pi r_H^3}-\frac{Q^2}{2\pi
r_H^5}-\frac{T}{2r_H^2})dr_H.
\end{eqnarray}
Thus it is not arbitrary to choose the step
length of $\triangle p$, it should be chosen
following the step length of $\triangle r_H$.

From the frequencies of small black holes shown in Tables
\ref{table3} and \ref{table31}, we can get the derivative of
$\omega$ with respect to $r_H$ at $r_H=1.35$ and $P=0.0002$,
$\partial_{r_H} \omega|_{r_H=1.35,P=0.002}\simeq-0.0007-0.0245I$ and
the derivative of $\omega$ with respect to $P$, $\partial_{P}
\omega|_{r_H=1.35,P=0.002}\simeq 62-41I$. Employing
Eq.(\ref{series}), we can estimate the quasinormal frequencies,
$\tilde{\omega}$, for small black holes from analytic linear
expansion. The analytically obtained $\tilde{\omega}$ are listed in
Table \ref{tabledb}, whose behaviors are in good agreement with the
numerical computation results listed in Table II with the increase
of the black hole size and the decrease of the pressure for small
black holes. Comparing $\triangle_1$ and $\triangle_2$ in Table
\ref{tabledb}, we see that for small black holes, the change of $P$
(or $l$) clearly wins over the change of the black hole size, which
dominantly contributes to the behavior of quasinormal frequencies
for small black hole phase.
\begin{center}
\begin{table}[ht]\label{tabledb}
\centering
\begin{tabular}{|c|c|c|c|c|}
\hline
$r_H$ &$P(10^{-3})$ &$\tilde{\omega}$ &$\triangle_1$&$\triangle_2$ \\[0.8ex] \hline
1.35&2.0& 0.2955-0.06489I& 0& 0\\[0.8ex] \hline
1.359&1.9 & 0.2893-0.06100I&0-0.00022I&-0.0062+0.0041I\\[0.8ex] \hline
1.368&1.8 & 0.2831-0.05713I&-0.000013-0.00044I&-0.012+0.0082I\\[0.8ex] \hline
1.377&1.7 & 0.2769-0.05325I&-0.000019-0.00066I&-0.019+0.012I\\[0.8ex] \hline
\end{tabular}
\caption{\label{tabledb} For the small black hole
phase, $\tilde{\omega}$ is the quasinormal
frequency from the linear approximation.
$\triangle_1$ and $\triangle_2$ are contributions
from changes of the black hole size and pressure,
respectively.}
\end{table}
\end{center}

For the large black hole phase, to keep the
linear approximation, we consider the small
change of the black hole size. Near  $P=0.0014$
and $r_H=7.880$ we have the derivatives
$\partial_{r_H}
\omega|_{P=0.0014,r_H=7.880}\simeq 0.012-0.032I$
and $\partial_{P}
\omega|_{P=0.0014,r_H=7.880}\simeq 135-178I $.
Using Eq.(\ref{series}), we can estimate the
quasinormal frequencies $\tilde{\omega}$ from the
linear approximation for the large black hole
phase as shown in Table \ref{tabledb1}. To keep
the linear approximation valid in
Eq.(\ref{series}), we just did the estimation in
the narrow range of $r_H$. $\tilde{\omega}$ keeps
the property of the quasinormal frequencies with
the increase of the black hole size and the
decrease of the pressure as listed in Table
\ref{table2} and Fig.\ref{arrow34} for large
black hole phase. $\triangle_1$ and $\triangle_2$
mark the contributions from the change of the
black hole size and the pressure, respectively.
It is clear that for the large black hole case,
the change of the horizon size contributes more
to the imaginary part of the frequency change.
Thus the perturbation decays faster following the
increase of the black hole size and has little
dependence on the change of the pressure or the
AdS length. The contributions of $\triangle_1$
and $\triangle_2$ on the real part of the
frequency are comparable. But the change of $P$
(or $l$) wins out a little. Thus with the
decrease of the $P$, the perturbation oscillation
becomes a bit quieter.

We would like to point out that the quasinormal
modes we considered are near the critical points
of the phase transition. In the approximation we
see that if the horizon $r_H$ becomes very big,
Eq.(\ref{totaldev}) will reduce to
$dP=-\frac{T}{2r_H^2}dr_H$,  so that $\triangle
P$ will be small enough and $\omega(r_H,P)$ will
slowly vary with $P$, going back to the behavior
in Table \ref{table31}. Thus only near the
critical point of the phase transition, we can
have the behavior of slopes of quasinormal modes
shown in Fig. 5, which is consistent with the
description in the linear approximation.

\begin{center}
\begin{table}[ht]\label{tabledb1}
\centering
\begin{tabular}{|c|c|c|c|c|}
\hline
$r_H$ &$P(10^{-3})$ &$\tilde{\omega}$ &$\triangle_1$&$\triangle_2$ \\[0.8ex] \hline
7.389&1.45& 0.2856-0.2405I& -0.0059+0.016& 0.0068-0.0089I\\[0.8ex] \hline
7.880&1.4 & 0.2848-0.2472I&0&0\\[0.8ex] \hline
8.388&1.35 & 0.2841-0.2544I&0.0061-0.016I&-0.0068+0.0089I\\[0.8ex] \hline
\end{tabular}
\caption{\label{tabledb1}For the large black hole
phase, $\tilde{\omega}$ is the quasinormal
frequency from the linear approximation.
$\triangle_1$ and $\triangle_2$ are contributions
from the changes of the black hole size and
pressure, respectively.}
\end{table}
\end{center}

\section{The behavior of quasinormal frequencies at the critical point}

For the isothermal phase transition, the critical point in Fig.(\ref{PT0})
appears when the isotherm system starts to have an inflection point at $P=P(r_H)$, which is given by
\begin{eqnarray}\label{isothermalcon}
\frac{\partial P}{\partial r_H}\Big|_{T=T_c, r_H=r_c}
=\frac{\partial^2 P}{\partial r_H^2}\Big|_{T=T_c, r_H=r_c}=0,
\end{eqnarray}
where $T_c=\frac{\sqrt{6}}{18\pi Q }$,
$r_c=\sqrt{6}Q$ and $P_c=\frac{1}{96\pi Q^2}$.
At $T=T_c$, a second-order phase transition occurs\cite{Gunasekaran}.
Fig.(\ref{critical-p-rh}) shows the $P-r_H$ diagram for fixed temperature. The line $1$ is for
$T> T_c$, where there is no phase transition in the system. Line $2$ marks the inflection point indicating the critical isotherm state $T=T_c$. Lines $3$ and $4$ are of $T<T_c$. The cross point of the dashed line shows the small-large black hole phase transition point at the $T=T_c$.
\begin{figure}[ht]\label{critical-p-rh}
\centering
\includegraphics[width=180pt]{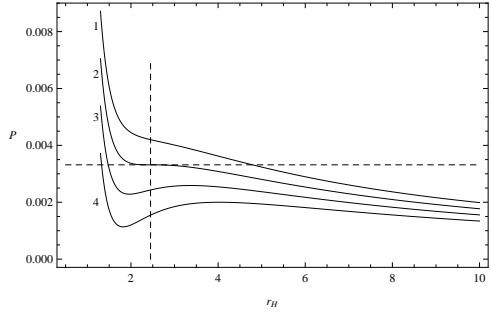}
\caption{\label{critical-p-rh} Lines marked 1-4 from the top to the bottom indicate the decrease of the temperature. Line 2
is for $T=T_c$. The cross of the dashed lines indicates the phase transition point of the critical isotherm state $T=T_c$.}
\end{figure}
Fig.(\ref{freng-crtical-Gp}) depicts the free energy at $T=T_c$, the cross of the dashed
lines indicates the phase transition point. The free energy continuously varies from the small black hole
phase to the large black hole phase which shows the character of a second-order phase transition.
The quasinormal frequencies of massless scalar perturbation for
different size black hole near the critical isothermal phase
transition point with $T=T_c$ are given in Table.\ref{tabletcdt}. The data above the horizontal line are the frequencies for small black holes, while those below are for large black holes. The quasinormal frequencies for small and large black holes keep the same behavior as the black hole horizon increases.

\begin{figure}[ht]\label{freng-crtical-Gp}
\centering
\includegraphics[width=180pt]{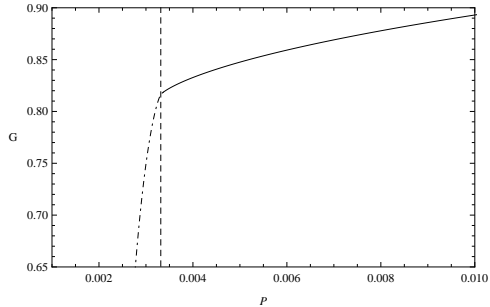}
\caption{\label{freng-crtical-Gp} The free energy of the critical isotherm $T=T_c$. The dashed line
indicates the phase transition point.}
\end{figure}

\begin{center}
\begin{table}[ht]\label{tabletcdt}
\centering
\begin{tabular}{|c|c|c|}
\hline
$P(10^{-3})$ & $r_{H}$ & $\omega$ \\[0.8ex] \hline
3.3160&2.376&0.4986-0.49766I  \\
3.3158&2.401&0.4982-0.49768I  \\
3.3157&2.425&0.4979-0.49770I  \\
3.3157&2.449&0.4976-0.49773I  \\\hline
3.3157&2.474&0.4973-0.49776I \\
3.3157&2.498&0.4969-0.49780I \\
3.3153&2.572&0.4959-0.49794I \\
3.3125&2.694&0.4942-0.49826I \\
\hline
\end{tabular}
\caption{\label{tabletcdt}. The quasinormal frequencies of small and large black holes in the
critical isothermal phase transition $T=T_c$. The data above the horizontal line are the
frequencies for small black holes, while those below are for large black holes.}
\end{table}
\end{center}

For the isobaric phase transition $T=T(r_H)$, the phase transition begins to happen at the critical position,
\begin{eqnarray}
\frac{\partial T}{\partial r_H}\Big|_{P=P_c, r_H=r_c}
=\frac{\partial^2 T}{\partial r_H^2}\Big|_{P=P_c, r_H=r_c}=0.
\end{eqnarray}
Fig.(\ref{critical-T-rh}) depicts the $T-r_H$ diagram for fixed $P$. Lines from top to bottom correspond to the decrease of the pressure. Line $2$ is for
$P=P_c$. The cross of the dashed lines indicates the small-large black hole phase transition point in the isobaric phase transition at the critical point. The free energy at the critical point is depicted in Fig.(\ref{freng-crtical-GT}), where the  dashed line marks the transition point. Table.\ref{tabletcdp} shows the quasinormal frequencies of the small and large black holes. The data above the horizontal
line are the frequencies for small black holes, while those below are for large black holes. From the
table we can see that at the critical point both the small and large black holes' quasinormal frequencies have the same
behavior as the black hole horizon increases.

From both the critical isothermal and isobaric phase transitions, we learn that at the critical point the quasinormal frequencies keep the same behavior as the black hole horizon increases. Below these critical points($P<P_c$,$T<T_c$), quasinormal frequencies can reflect the phase transition between small and large black holes. Due to our numerical code efficiency, the difference in quasinormal frequencies can be clearer in a state much below the critical point.

\begin{figure}[ht]\label{critical-T-rh}
\centering
\includegraphics[width=180pt]{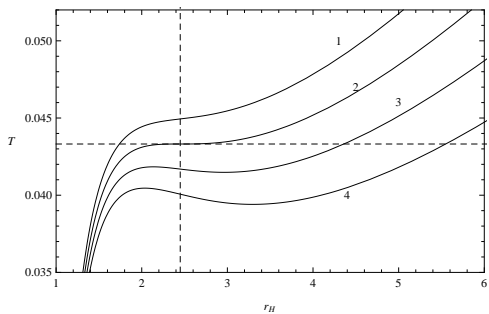}
\caption{\label{critical-T-rh} The lines from top to bottom are for different fixed pressure. Line 2 is for $P=P_c$. The cross point of the dashed lines indicates the phase transition point of the isobaric system at the critical point $P=P_c$.}
\end{figure}

\begin{figure}[ht]\label{freng-crtical-GT}
\centering
\includegraphics[width=180pt]{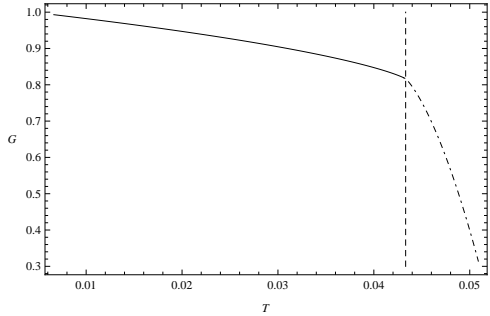}
\caption{\label{freng-crtical-GT}The free energy of the isobaric system at the critical point $P=P_c$. The dashed line
indicates the phase transition point. }
\end{figure}

\begin{center}
\begin{table}[ht]\label{tabletcdp}
\centering
\begin{tabular}{|c|c|c|}
\hline
$T(10^{-2})$ & $r_{H}$ & $\omega$ \\[0.8ex] \hline
4.32&2.079&0.5006-0.4968I  \\
4.32&2.103&0.5006-0.4969I  \\
4.33&2.168&0.5005-0.4973I  \\
4.33&2.449&0.4976-0.4977I  \\\hline
4.34&2.790&0.4933-0.4991I  \\
4.34&2.889&0.4924-0.5000I \\
4.34&2.962&0.4919-0.5008I \\
4.35&3.022&0.4916-0.5016I \\
\hline
\end{tabular}
\caption{\label{tabletcdp} The quasinormal frequencies of small and large black holes in the isobaric phase transition at $P=P_c$. The data above the horizontal line are the
frequencies for small black holes, while those below are for large black holes.}
\end{table}
\end{center}

\section{Conclusion}
\label{4s}

We have calculated the quasinormal modes  of massless scalar
perturbations around small and large four-dimensional RN-AdS black
holes. We found that when the Van der Waals analogy thermodynamic
phase transition happens, no matter in the isobaric process by
fixing the pressure in the extended space or in the isothermal
process by fixing the temperature of the system, the slopes of the
quasinormal frequency change drastically different in the small and
large black holes as we increase the value of the horizon radius.
This clearly presents the signature of the phase transition between
small and large black holes. This is one more example shows that the
quasinormal mode can provide dynamical physical phenomenon of the
thermodynamic phase transition. Since the quasinormal mode is
expected to be detected and has strong astrophysical interest, its
ability to reflect the thermodynamic phase transition is
interesting, which is expected to disclose the observational
signature of the thermodynamic phase transition.

{\bf Acknowledgments}

This work was supported by the National Natural Science Foundation of China.
Yunqi Liu was also supported by the China Postdoctoral Science Foundation.


\begin{thebibliography}{99}

\bibitem{HawkingPage83}
S. W. Hawking and D.N. Page, \emph{Thermodynamics of black holes in
anti-de Sitter space}, Comm. Math. Phys. \textbf{87} (1983) 577.

\bibitem{Chamblin99}
 A. Chamblin, R. Emparan, C. V. Johnson and R. C. Myers,
 \emph{Charged AdS Black Holes and Catastrophic Holography}, Phys. Rev. D \textbf{60}, 064018 (1999), [hep-th/9902170].

\bibitem{Chamblin992}
A. Chamblin, R. Emparan, C. V. Johnson and R. C. Myers,
\emph{Holography, Thermodynamics and Fluctuations of Charged AdS
Black Holes}, Phys. Rev. D \textbf{60}, 104026 (1999),
[hep-th/9904197].

\bibitem{Niu12}
 C. Niu, Y. Tian and X. N. Wu, \emph{Critical Phenomena and Thermodynamic Geometry of RN-AdS Black Holes},
 Phys. Rev. D \textbf{85}, 024017 (2012), [arXiv:1104.3066].

\bibitem{Fernando06}
S. Fernando, \emph{Thermodynamics of Born-Infeld-anti-de Sitter
black holes in the grand canonical ensemble}, Phys. Rev. D
\textbf{74}, 104032 (2006), [hep-th/0608040].

\bibitem{Dey07}
T. K. Dey, S. Mukherji, S. Mukhopadhyay and S. Sarkar, \emph{Phase
Transitions in Higher Derivative Gravity}, JHEP \textbf{0704}, 014
(2007), [hep-th/0609038].

\bibitem{Dey072}
T. K. Dey, S. Mukherji, S. Mukhopadhyay and S. Sarkar, \emph{Phase
transitions in higher derivative gravity and gauge theory: R-charged
black holes}, JHEP \textbf{0709}, 026 (2007), [arXiv:0706.3996].

\bibitem{Anninos}
D. Anninos and G. Pastras, \emph{Thermodynamics of the
Maxwell-Gauss-Bonnet anti-de Sitter Black Hole with Higher
Derivative Gauge Corrections}, JHEP \textbf{0907}, 030 (2009),
[arXiv:0807.3478].

\bibitem{Poshteh}
M. B. J. Poshteh, B. Mirza and Z. Sherkatghanad, \emph{Phase
transition, critical behavior, and critical exponents of Myers-Perry
black holes}, Phys. Rev. D \textbf{88}, 024005 (2013),
[arXiv:1306.4516].

\bibitem{WeiLiu}
S. W. Wei and Y. X. Liu, \emph{Critical phenomena and thermodynamic
geometry of charged Gauss-Bonnet AdS black holes}, Phys. Rev. D
\textbf{87}, 044014 (2013), [arXiv:1209.1707].

\bibitem{Lala}
A. Lala, \emph{Critical phenomena in higher curvature charged AdS
black holes}, Advances in High Energy Physics, Volume 2013, Article
ID 918490, [arXiv:1205.6121].

\bibitem{Tsai}
Y. D. Tsai, X. N. Wu and Y. Yang, \emph{Phase Structure of Kerr-AdS
Black Hole}, Phys. Rev. D \textbf{85}, 044005 (2012),
[arXiv:1104.0502].

\bibitem{Banerjee1}
R. Banerjee and D. Roychowdhury, \emph{Critical behavior of Born
Infeld AdS black holes in higher dimensions}, Phys. Rev. D
\textbf{85}, 104043 (2012), [arXiv:1203.0118].

\bibitem{Banerjee2}
R. Banerjee and D. Roychowdhury, \emph{Thermodynamics of phase
transition in higher dimensional AdS black holes}, JHEP
\textbf{1111}, 004 (2011), [arXiv:1109.2433].

\bibitem{Banerjee3}
R. Banerjee, S. K. Modak and S. Samanta, \emph{Second Order Phase
Transition and Thermodynamic Geometry in Kerr-AdS Black Hole}, Phys.
Rev. D \textbf{84}, 064024 (2011), [arXiv:1005.4832].

\bibitem{Banerjee4}
R. Banerjee, S. K. Modak and S. Samanta, \emph{Glassy Phase
Transition and Stability in Black Holes}, Eur. Phys. J. C
\textbf{70}, 317 (2010), [arXiv:1002.0466].

\bibitem{Banerjee11}
R. Banerjee, S. Ghosh and D. Roychowdhury, \emph{New type of phase
transition in Reissner Nordstrom-AdS black hole and its
thermodynamic geometry}, Phys. Lett. B \textbf{696}, 156 (2011),
[arXiv:1008.2644].

\bibitem{Dolan}
B. P. Dolan, \emph{Pressure and volume in the first law of black
hole thermodynamics}, Class. Quant. Grav. \textbf{28}, 235017 (2011)
, [arXiv:1106.6260].

\bibitem{Dolan2}
B. P. Dolan, \emph{The cosmological constant and the black hole
equation of state}, Class. Quant. Grav. \textbf{28}, 125020 (2011),
[arXiv:1008.5023].

\bibitem{Dolan3}
B. P. Dolan,\emph{ Compressibility of rotating black holes}, Phys.
Rev. D \textbf{84}, 127503 (2011), [arXiv:1109.0198].

\bibitem{DolanKastor}
B. P. Dolan, D. Kastor, D. Kubiznak, R. B. Mann and J. Traschen,
\emph{Thermodynamic Volumes and Isoperimetric Inequalities for de
Sitter Black Holes}, Phys. Rev. D \textbf{87}, 104017 (2013),
[arXiv:1301.5926].

\bibitem{Spallucci}
E. Spallucci and A. Smailagic, \emph{Maxwell's equal area law for
charged Anti-deSitter black holes}, Phys. Lett. B 723, 436 (2013),
[arXiv:1305.3379].


\bibitem{Kastor:2009wy}
D. Kastor, S. Ray and J. Traschen, \emph{Enthalpy and the Mechanics
of AdS Black Holes}, Class. Quant. Grav. {\bf 26}, 195011 (2009),
[arXiv:0904.2765].


\bibitem{Kubiznak}
D. Kubiznak and R. B. Mann, \emph{P-V criticality of charged AdS
black holes}, JHEP 1207, 033 (2012), [arXiv:1205.0559].

\bibitem{Gunasekaran}
S. Gunasekaran, R. B. Mann and D. Kubiznak, \emph{Extended phase
space thermodynamics for charged and rotating black holes and
Born-Infeld vacuum polarization}, JHEP \textbf{1211}, 110 (2012),
[arXiv:1208.6251].

\bibitem{BelhajChabab}
A. Belhaj, M. Chabab, H. ElMoumni and M. B. Sedra, \emph{On
Thermodynamics of AdS Black Holes in Arbitrary Dimensions}, Chin.
Phys. Lett. \textbf{29}, 100401 (2012), [arXiv:1210.4617];

\bibitem{Hendi}
S. H. Hendi and M. H. Vahidinia, \emph{Extended phase space
thermodynamics and P-V criticality of black holes with nonlinear
source}, Phys. Rev. D \textbf{88}, 084045 (2013), [arXiv:1212.6128].

\bibitem{BelhajChabab1}
A. Belhaj, M. Chabab, H. E. Moumni, L. Medari and M. B. Sedra,
\emph{On Thermodynamical Behaviors of Kerr-Newman AdS Black Holes},
Chin. Phys. Lett. \textbf{30}, 090402 (2013), [arXiv:1307.7421].

\bibitem{Belhaj}
A. Belhaj, M. Chabab, H. E. Moumni and M. B. Sedra, \emph{Critical
Behaviors of 3D Black Holes with a Scalar Hair}, [arXiv:1306.2518].

\bibitem{ChenLiu}
S. Chen, X. Liu, C. Liu and J. Jing, \emph{P-V criticality of AdS
black hole in f(R) gravity}, Chin. Phys. Lett. \textbf{30}, 060401
(2013), [arXiv:1301.3234].

\bibitem{Altamirano:2013ane}
N. Altamirano, D. Kubiznak and R. B. Mann, \emph{Reentrant Phase
Transitions in Rotating AdS Black Holes}, Phys. Rev. D {\bf 88},
101502 (2013), [arXiv:1306.5756].

\bibitem{Altamirano:2013uqa}
N. Altamirano, D. Kubiznak, R. B. Mann and Z. Sherkatghanad,
\emph{Kerr-AdS analogue of triple point and solid/liquid/gas phase
transition}, Class. Quant. Grav. {\bf 31}, 042001 (2014),
[arXiv:1308.2672].

\bibitem{CaiCao}
R. G. Cai, L. M. Cao, L. Li and R. Q. Yang, \emph{P-V criticality in
the extended phase space of Gauss-Bonnet black holes in AdS space},
JHEP \textbf{1309}, 005 (2013), [arXiv:1306.6233].

\bibitem{XuXu}
W. Xu, H. Xu and L. Zhao, \emph{Gauss-Bonnet coupling constant as a
free thermodynamical variable and the associated criticality},
[arXiv:1311.3053].

\bibitem{Dutta}
S. Dutta, A. Jain and R. Soni, \emph{Dyonic Black Hole and
Holography}, JHEP \textbf{1312}, 060 (2013), [arXiv:1310.1748].

\bibitem{Zou:2013owa}
D. C.Zou, S. J. Zhang and B. Wang, \emph{Critical behavior of
Born-Infeld AdS black holes in the extended phase space
thermodynamics}, Phys. Rev. D {\bf 89}, 044002 (2014),
[arXiv:1311.7299].

\bibitem{Mo:2014mba}
J. X.~Mo and W. B. Liu, \emph{Ehrenfest scheme for P-V criticality
of higher dimensional charged black holes, rotating black holes and
Gauss-Bonnet AdS black holes}, Phys. Rev. D {\bf 89}, 084057 (2014),
[arXiv:1404.3872].

\bibitem{Zou:2014mha}
D.~C.~Zou, Y.~Liu and B.~Wang, \emph{Critical behavior of charged
Gauss-Bonnet AdS black holes in the grand canonical ensemble},
[arXiv:1404.5194].

\bibitem{Zhao}
R. Zhao, H. H. Zhao, M. S. Ma and L. C. Zhang, \emph{On the critical
phenomena and thermodynamics of charged topological dilaton AdS
black holes},Eur. Phys. J. C \textbf{73 }(2013) 2645,
[arXiv:1305.3725].

\bibitem{Nollert}
H. P. Nollert, \emph{Quasinormal modes: the characteristic sound of
black holes and neutron stars}, Class. Quant. Grav. \textbf{16},
R159 (1999).

\bibitem{Kokkotas}
K. D. Kokkotas and B. G. Schmidt, \emph{Quasi-normal modes of stars
and black holes}, Living Rev. Rel. 2, 2 (1999).

\bibitem{WangBraz}
B. Wang, \emph{Perturbations around black holes}, Braz. J. Phys.
\textbf{35}, 1029 (2005).

\bibitem{KonoplyaZhidenko}
R. A. Konoplya and A. Zhidenko, \emph{Quasinormal modes of black
holes: from astrophysics to string theory}, Rev. Mod. Phys. 83, 793
(2011), [arXiv:1102.4014].

\bibitem{Davies}
P. C. W. Davies, \emph{The thermodynamic theory of black holes},
Proc. Roy. Soc. Lond. A 353, 499 (1977).

\bibitem{Davies2}
P. C. W. Davies, \emph{Thermodynamic phase transitions of
Kerr-Newman black holes in de Sitter space}, Class. Quant. Grav.
\textbf{6}, 1909 (1989).

\bibitem{Jingpan08}
J. Jing and Q. Pan, \emph{Quasinormal modes and second order
thermodynamic phase transition for Reissner-Nordstrom black hole},
Phys. Lett. B \textbf{660}, 13 (2008).

\bibitem{BertiCardoso}
E. Berti, V. Cardoso, \emph{Quasinormal modes and thermodynamic
phase transitions}, Phys. Rev. D \textbf{77}, 087501 (2008).

\bibitem{He:2008im}
X.~He, B.~Wang, S.~Chen, R.~G.~Cai and C.~Y.~Lin, \emph{Quasinormal
modes in the background of charged Kaluza-Klein black hole with
squashed horizons}, Phys. Lett. B {\bf665}, 392 (2008)
[arXiv:0802.2449].



\bibitem{Gubser:2000mm}
S.~S.~Gubser and I.~Mitra, \emph{The evolution of unstable black
holes in anti-de Sitter space}, JHEP {\bf 0108}, 018 (2001),
[hep-th/0011127].

\bibitem{KoutsoumbasMusiri}
G. Koutsoumbas, S. Musiri, E. Papantonopoulos, G. Siopsis,
\emph{Quasi-normal Modes of Electromagnetic Perturbations of
Four-Dimensional Topological Black Holes with Scalar Hair}, JHEP
\textbf{0610} (2006) 006.

\bibitem{ShenWang}
J. Shen, B. Wang, C. Y. Lin, R. G. Cai and R. K. Su, \emph{The phase
transition and the Quasi-Normal Modes of black Holes}, JHEP
\textbf{0707}, 037 (2007).

\bibitem{Koutsoumbas:2008pw}
G.~Koutsoumbas, E.~Papantonopoulos and G.~Siopsis, \emph{Phase
Transitions in Charged Topological-AdS Black Holes}, JHEP {\bf
0805}, 107 (2008), [arXiv:0801.4921].

\bibitem{Rao:2007zzb}
X. P.~Rao, B.~Wang and G. H.~Yang, \emph{Quasinormal modes and phase
Transition of black holes}, Phys. Lett. B {\bf 649}, 472 (2007),
[arXiv:0712.0645].

\bibitem{Myung:2008ze}
Y.~S.~Myung, \emph{Phase transition for black holes with scalar hair
and topological black holes}, Phys. Lett. B {\bf 663}, 111 (2008),
[arXiv:0801.2434].

\bibitem{Gubser:2000ec}
S.~S.~Gubser and I.~Mitra, \emph{Instability of charged black holes
in anti-de Sitter space}, [hep-th/0009126].

\bibitem{He:2010zb}
X.~He, B.~Wang, R.~G.~Cai and C.~Y.~Lin, \emph{Signature of the
black hole phase transition in quasinormal modes}, Phys. Lett. B
{\bf 688}, 230 (2010), [arXiv:1002.2679].

\bibitem{Cai:2011qm}
R.~G.~Cai, X.~He, H.~F.~Li and H.~Q.~Zhang, \emph{Phase transitions
in AdS soliton spacetime through marginally stable modes}, Phys.
Rev. D {\bf 84}, 046001 (2011), [arXiv:1105.5000].

\bibitem{Liu:2011cu}
Y.~Liu and B.~Wang, \emph{Perturbations around the AdS Born-Infeld
black holes}, Phys. Rev. D {\bf 85}, 046011 (2012),
[arXiv:1111.6729].

\bibitem{Abdalla:2010nq}
E.~Abdalla, C.~E.~Pellicer, J.~de Oliveira and A.~B.~Pavan,
\emph{Phase transitions and regions of stability in
Reissner-Nordstrom holographic superconductors}, Phys. Rev. D {\bf
82}, 124033 (2010), [arXiv:1010.2806].

\bibitem{zhuWang}
J. M. Zhu, B. Wang, and E. Abdalla, \emph{Object picture of
quasinormal ringing on the background of small Schwarzschild Anti-de
Sitter black holes}, Phys. Rev. D \textbf{63}, 124004 (2001),
[hep-th/0101133].

\bibitem{WangMolina}
B. Wang, C. Molina, and E. Abdalla, \emph{Evolving of a massless
scalar field in Reissner-Nordstrom Anti-de Sitter spacetimes}, Phys.
Rev. D \textbf{63}, 084001 (2001), [hep-th/0005143].

\end{thebibliography}
\end{document}